\documentclass[11pt,twocolumn,twoside,a4paper,amsmath,amssymb,aps,showkeys,showpacs]{revtex4}

\usepackage{amsfonts}
\usepackage{graphics} 
\usepackage{epsfig}
\usepackage{fancyheadings}

\begin{document}
\thispagestyle{myheadings}
\rhead[]{}
\lhead[]{}
\chead[G.Wilk, Z.W\l odarczyk]{Multiplicity fluctuations and temperature fluctuations}

\title{Multiplicity fluctuations and temperature fluctuations}

\author{Grzegorz Wilk}
\email{wilk@fuw.edu.pl}

\affiliation{%
The Andrzej So{\l}tan Institute for Nuclear Studies, Ho\.{z}a 69,
PL-00681, Warsaw, Poland}%

\author{Zbigniew W\l odarczyk}
\email{wlod@ujk.kielce.pl}

\affiliation{Institute of Physics, Jan Kochanowski University,
             \'Swi\c{e}tokrzyska 15, PL-25406 Kielce, Poland  }

\received{ ?????????? }

\begin{abstract}
We argue that specific fluctuations observed in high-energy
nuclear collisions can be attributed to intrinsic fluctuations of
temperature of the hadronizing system formed in such processes and
therefore can be described by the same nonextensivity parameter
$q$ characterizing Tsallis statistics describing such systems (for
$q \rightarrow 1$ one recovers the usual Boltzmann-Gibbs
approach). It means that $|q-1|$, which is a direct measure of
temperature fluctuations, can also be expressed by the observed
mean multiplicity, $\langle N\rangle$, and by its variance,
$Var(N)$. This allows to deduce from the experimental data the
system size dependence of parameter $q$ with $q = 1$ corresponding
to an infinite, thermalized source with a fixed temperature, and
with the observed $q > 1$ corresponding to a finite source in
which both the temperature and energy fluctuate.
\end{abstract}



\maketitle

\renewcommand{\thefootnote}{\fnsymbol{footnote}}

\renewcommand{\thefootnote}{\roman{footnote}}


Since already some time we advocate that most of the single
particle distributions measured in high energy collisions when
analyzed by statistical hadronization model should use its
nonextensive version (for all details concerning nonextensive
statistical physics and its applications see recent review
\cite{T}, our works are summarized there in \cite{EPJA}). In such
approach a standard Boltzmann-Gibbs (BG) exponential
distributions, $g(E) = C\exp( - E/T)$, are replaced by Tsallis
distribution ($q$-exponential), $h_q(E) = C_q\left[ 1 - (1-q)E/T
\right]^{1/(1-q)}$, and a new phenomenological parameter $q$
occurs which accounts summarily for the possible intrinsic
fluctuations of the usual parameter, the temperature $T$ of the
hadronizing system, $q - 1 = Var(T)/\langle T\rangle ^2$ (for $q
\rightarrow 1$ one recovers the usual BG approach). Because
$Var(T)/\langle T\rangle ^2 = 1/C_V$ ($C_V$ is the heat capacity
under constant volume) it means that $q - 1 = 1/C_V$ in fact
measures the heat capacity of hadronizing system. In nuclear
collisions in which only part of nucleus participates actively in
the collision, the above picture should be slightly modified by
allowing for the energy transfer to/from the collision region to
its surroundings (formed by the spectator nucleons), As result $T$
in the above formulas is replaced by $T_{eff} = T_0 +
(q-1)T_{visc}$, with $T_0 = \langle T\rangle$ and $T_{visc}$ being
a new parameter characterizing this energy transfer
\cite{EPJA,OurPaper}. On the other hand, it was also shown in
\cite{EPJA} that whereas independently produced particles
following BG distributions in energy show Poisson multiplicity
distributions, the same particles following Tsallis distributions
instead are inevitably distributed according to the Negative
Binomial formula, $P(N) = \Gamma(N+k)/\left[ \Gamma(N+1)\Gamma(k)
\right] = \left(\langle N\rangle \right)^N/\left( 1 + \langle
N\rangle/k\right)^{N+1}$ with $k = 1/(q-1)$ (or $q = 1 + 1/k$) and
$Var(N) = \langle N\rangle + (q-1)\langle N\rangle^2$.

Let us now observe that for a system with finite size remaining in
contact with a heat bath one has, following Lindhard's approach
\cite{Lin}, that
\begin{equation}
Var(U) + C_V^2 Var(T) = \langle T\rangle ^2 C_V. \label{eq:unrel}
\end{equation}
This is a kind of uncertainty relation (in the sense that it
expresses the truth that in the case of conjugate variables one
standard deviation in some measurement can only become small at
the expense of the increase of some other standard deviation
\cite{UnRel}). Relation (\ref{eq:unrel}) is supposed to be valid
all the way from the {\it canonical ensemble},
\begin{equation}
Var(T) = 0\quad {\rm and}\quad Var(U) = \langle T\rangle ^2 C_V,
\label{eq:canonical}
\end{equation}
to the {\it microcanonical ensemble},
\begin{equation}
Var(T) = \langle T\rangle ^2/C_V\quad {\rm and}\quad  Var(U) =
0.\label{eq:microcanonical}
\end{equation}
It means therefore that Eq. (\ref{eq:unrel}) expresses both the
complementarity between the temperature and energy, and between
the canonical and microcanonical description of the system. This
should be understood in the same way as the (improper) eigenstates
of position and momentum appear as extreme cases in the quantum
mechanical uncertainty relations. It is worth knowing that in
\cite{limq} the limiting cases of Tsallis statistics was
investigated in which $q$ was interpreted as a measure thermal
bath heat capacity: $q = 1$, i.e. canonical, case would correspond
to an infinite bath (thermalized and with fixed temperature),
whereas $ q = - \infty $, i.e. microcanonical, case would
correspond to a bath with null heat capacity (isolated and with
fixed energy). All intermediate cases would then correspond to the
finite heat capacity (both temperature and energy fluctuate).

\begin{figure}[t]
   \includegraphics[width=7.5cm]{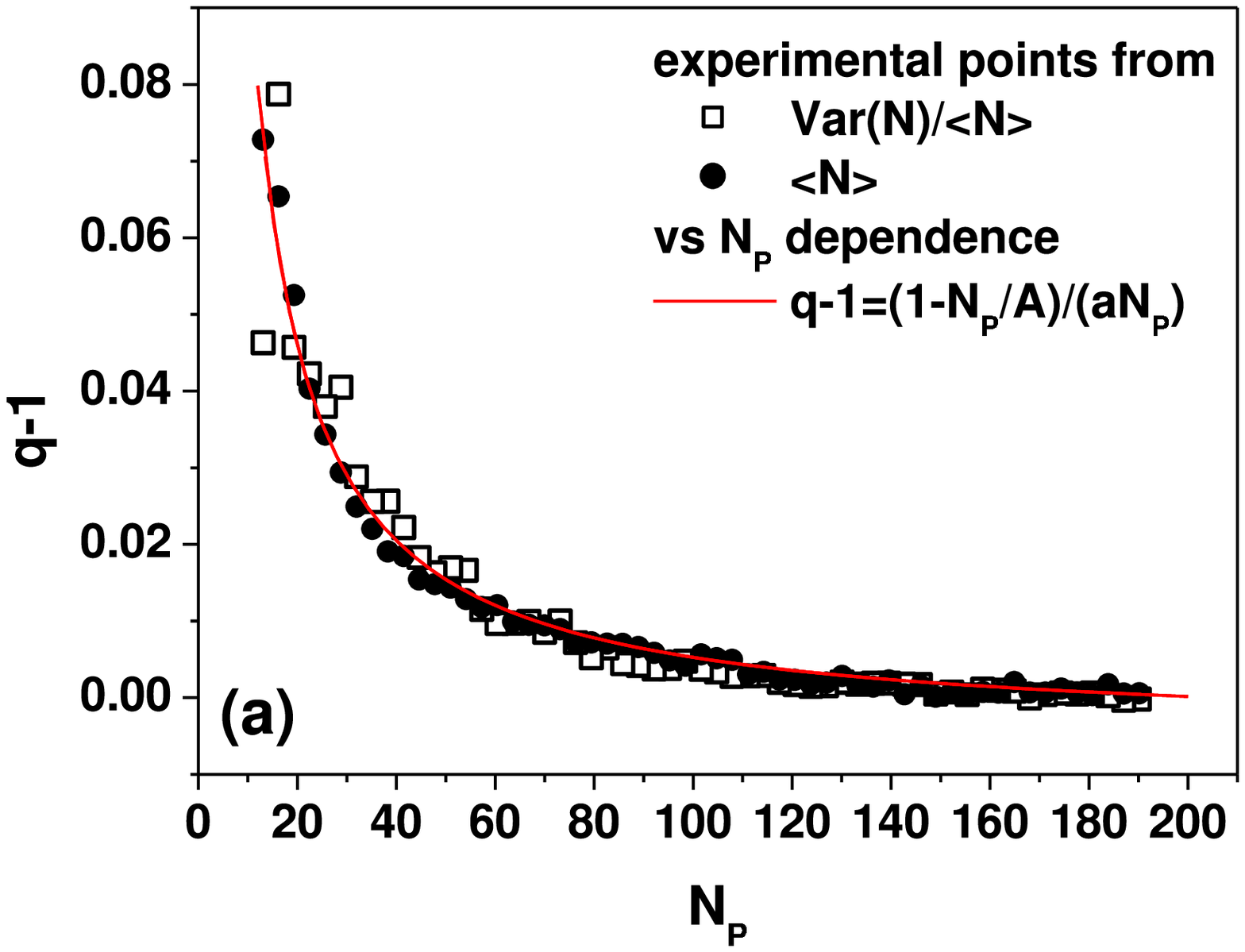}\\
   \includegraphics[width=7.5cm]{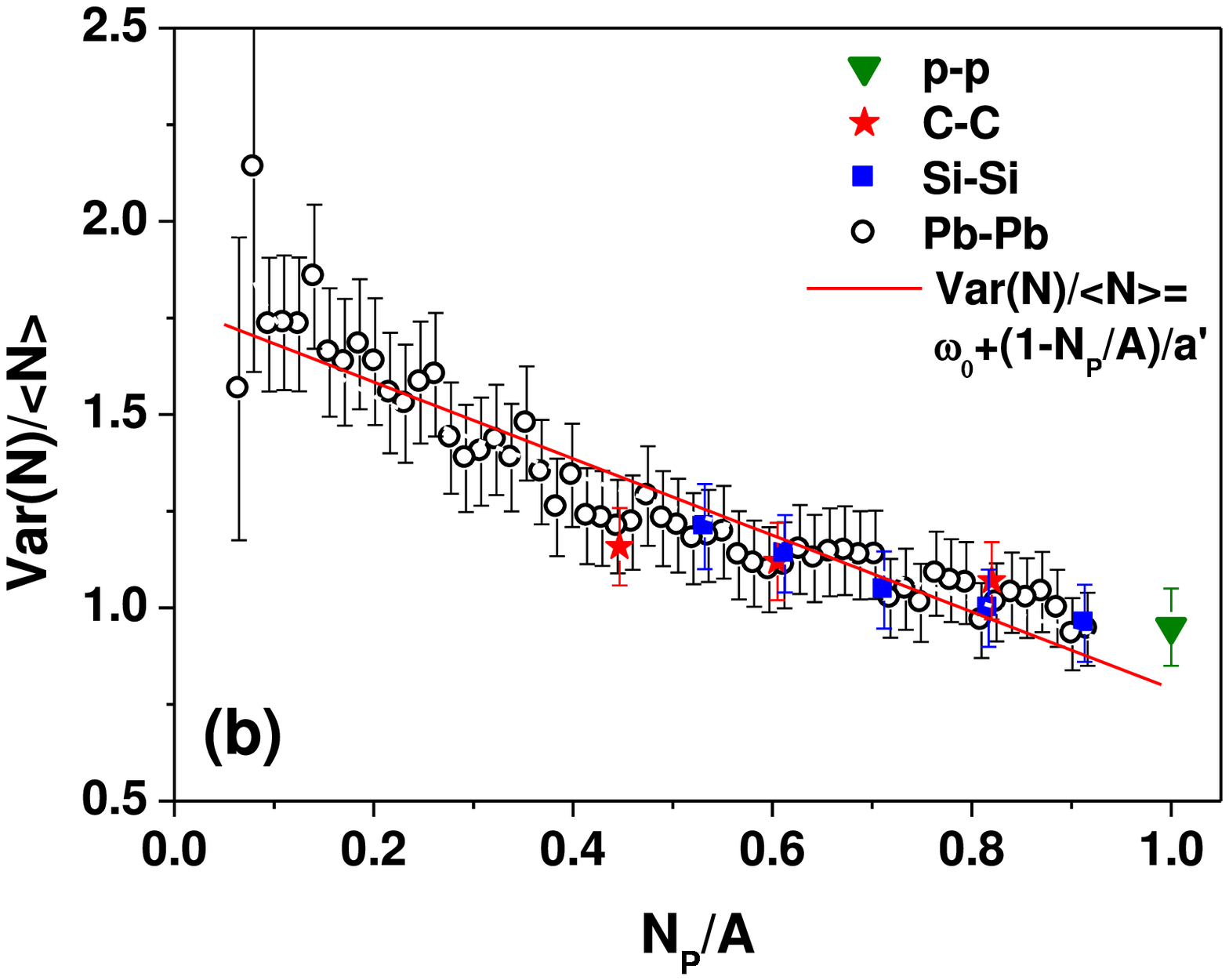}
   \caption{(Color on line). $(a)$ Prediction of Eq. (\ref{eq:q-1})
            versus experimental results obtained from $Var(N)/\langle N\rangle$
            (squares) and from the $\langle N\rangle$ (circles) - see \cite{OurPaper}
            for details.
            $(b)$ $Var(N)/\langle N\rangle$ for negatively
            charged particles produced in p+p, semicentral C+C, semicentral
            Si+Si, and Pb+Pb collisions as function of the fraction of nucleons
            participating in the collision, $N_P/A$, compared with our predictions
            (see \cite{OurPaper} for details).}
   \label{Fig:1}
\end{figure}

To obtain realistic (intermediate) distributions let us start from
a system at a fixed temperature $T$. The standard deviation of its
energy is
\begin{equation}
 Var(U) = \langle T\rangle^2
\frac{\partial \langle U\rangle}{\partial T} = \langle T\rangle ^2
C_V. \label{eq:eq1}
\end{equation}
Inverting the canonical distribution $g_T(U)$ one can obtain
\begin{equation} g_U(T) = -T^2 \frac{\partial}{\partial T}
\int^U_0g_T(U')dU' \label{eq:invert}
\end{equation}
and interpret it as a probability distribution of the temperature
in the system. The standard deviation of this distribution then
yields $Var(T) = \langle T\rangle ^2/C_V$. Because for a
canonically distributed system the energy variance is $ Var(U) =
\langle T\rangle^2 C_V$ and because for an isolated system $Var(U)
= 0$, for the intermediate case the variance (expressing energy
fluctuations in the system) can be assumed to be equal to
\begin{equation}
Var(U) = \langle T\rangle^2 C_V \xi,\quad \xi \in (0,1),
\label{eq:eq2}
\end{equation}
where the parameter $\xi$ depends on the size of the hadronizing
source. Inserting this in Eq. (\ref{eq:unrel}) one gets that $q$
depends on $\xi$ in the following way:
\begin{equation}
q - 1 = \frac{Var(T)}{\langle T\rangle^2} = \frac{1 - \xi}{C_V}.
\label{eq:relation}
\end{equation}
It is natural to assume that the size of the thermal system
produced in heavy ion collisions is proportional to the number of
participating nucleons $N_P$, $\xi \simeq N_P/A$. Because $C_V
\cong a N_P $, we obtain that
\begin{equation} q - 1 = \frac{1}{a N_P}\left( 1- \frac{N_P}{A} \right). \label{eq:q-1}
\end{equation}
As shown in detail in \cite{OurPaper}, this is precisely what is
observed experimentally (cf., \ref{Fig:1}a). To recapitulate what was done there:\\
$(i)$ We know that if $U =$ const and $T =$ const then the
multiplicity distribution $P(N)$ is Poissonian and we also know
how fluctuations of $T$ change $P(N)$ from the Poissonian form to
the NBD one.\\ $(ii)$ We want now to see how big are fluctuations
of $T$ in our hadronizing systems formed in a collision process.
It turns out that for $N =$ const we have either Eq.
(\ref{eq:eq1}) or Eq. (\ref{eq:microcanonical}) depending on
whether $T =$ const or $U =$ const. Assuming now validity of Eq.
(\ref{eq:unrel}) not only in the above limiting cases but also in
the general case when both the energy $U$ and the temperature $T$
fluctuate at the same time (fluctuations of the energy $U$ are
given by Eq. \ref{eq:eq2}) and corresponding fluctuations of the
temperature $T$ are given by Eq. (\ref{eq:relation})).\\ $(iii)$
Knowing how big are fluctuations of $T$ we can deduce fluctuations
of the multiplicity $N$.

The analysis of different aspects of such fluctuations was
performed in \cite{OurPaper} and will not be repeated here.
Instead we would like to bring ones attention to the following
remarks. In the above formalism we have (cf. Eq. (B3) of
\cite{OurPaper} ) that $g_{T,N}(U) = [\beta/\Gamma(N)](\beta
U)^{N-1}\exp( - \beta U)$ with fluctuations given by
$Var(U)/\langle U\rangle^2 = 1/N$. In fact one can now invert
distribution $g_{T,N}(U)$ (proceeding in analogous way as in
Appendix B of \cite{OurPaper}) to obtain multiplicity
distributions $g_{T,U}(N)$ and obtain distribution of temperature,
$g_{U,N}(T) = \frac{\partial}{\partial \beta}\int_0^U g_{T,N}(U')
dU' = [U/\Gamma(N)](\beta U)^{N-1} \exp(-\beta U)$ with
fluctuations $Var(T)/\langle T\rangle ^2 \simeq Var(\beta)/\langle
\beta \rangle ^2 = 1/N  $. The fluctuations of temperature
obtained this way have gamma distribution. Because $C_V \propto N$
one obtains Eqs. (\ref{eq:eq1}) and (\ref{eq:microcanonical}).
This illustrates that we can deduce fluctuations (i.e., the
corresponding probability distributions) of any quantity out of
($T, U, N$) provided only that the other two are constant. The
implication of this fact are still to be discussed.

Let us return again to results of \cite{OurPaper}. The Eq.
(\ref{eq:q-1}) can be translated to direct dependence of
$Var(N)/\langle N\rangle$ on the number of participants $N_P$, If
$T_{eff}$ depends on $q$, i.e., if $T_{visc} \neq 0$ and there is
some energy transfer from the hadronization region to the
spectators, then the above dependence is connected with the
dependence of $\langle N\rangle$ on $N_P$. The observed decreasing
of $Var(N)/\langle N\rangle$ with $N_P$ implies then the {\it
nonlinear} dependence of $\langle N\rangle$ on the number of
participants $N_P$. It was also shown in \cite{OurPaper} that
$Var(N)/\langle N\rangle \propto 1 - N_P/A$ (see Fig.
\ref{Fig:1}b). In fact, also transverse momentum fluctuations
defined by the measure $\Phi$ behave in the similar way: $\Phi
\propto 1 - N_P/A$. It means then that these fluctuations are
reflecting not so much fluctuations of transverse momenta but
rather fluctuations of the multiplicity.

We would like to add here a new observation concerning behavior of
correlations in transverse momenta measured in \cite{STAR}. As
shown in \cite{OurOld} one can connect measure of fluctuations
$\Phi$ with measured covariance in transverse momenta
$Cov(p_{Ti},p_{Tj})$:
\begin{eqnarray}
\Phi \left( \Phi\! +\! 2 \sqrt{Var\left( p_t \right)}\right)\!
&=&\! \left( \langle N\rangle\! -\! 1\! + Var(N)/\langle
N\rangle\right)\cdot \nonumber\\
&& \cdot  Cov\left( p_{Ti},p_{Tj}\right). \label{eq:CovPhi}
\end{eqnarray}
Because we observe that $\Phi << 2\sqrt{Var\left(p_T\right)}$, one
can write approximately that
\begin{equation}
\!\! \frac{Cov\left((p_{Ti},p_{Tj}\right)}{2\sqrt{Var(p_T)} }\!\!
\simeq \!\! \frac{\Phi}{\langle N\rangle\! -\! 1\! +
Var(N)/\langle N\rangle}\!=\!\frac{\Phi}{\langle N\rangle q}.
\label{eq:CovApprox}
\end{equation}
Using now relations derived in \cite{OurPaper} one gets that
\begin{eqnarray}
\frac{Cov\left((p_{Ti},p_{Tj}\right)}{2\sqrt{Var(p_T)} }\simeq
\frac{\Phi_{N_P=A} + b \left(1 - \frac{N_P}{A}\right)}{n_0N_P +
\frac{c+1}{a'}\left(1 - \frac{N_P}{A}\right)}.  \label{eq:final}
\end{eqnarray}
\begin{figure}[t]
   \includegraphics[width=7.5cm]{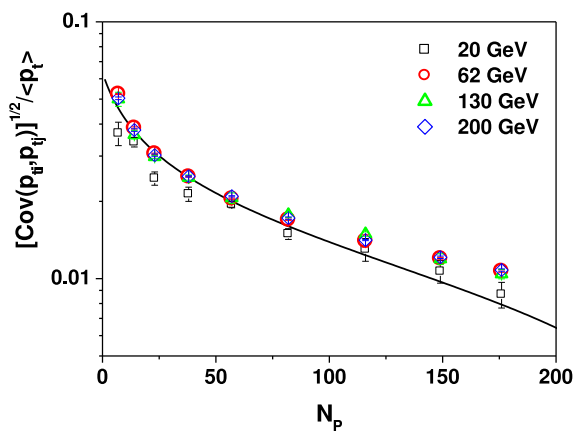}
   \vspace{-5mm}
   \caption{(Color on line). Our results for correlations.
   Data are from \cite{STAR}.
   The full line is our prediction as given by Eq.
   (\ref{eq:final}). See text for further details.}
   \label{Fig:2}
\end{figure}
This is our prediction and it is shown in Fig. \ref{Fig:2} for
$\Phi_{N_P=A} = 1.44$ MeV and for other parameters obtained from
different comparisons with multiplicity fluctuations done in
\cite{OurPaper}, namely: $\langle p_T\rangle = 400$ MeV, $b = 4.8$
MeV, $n_0 = 0.642$, $c = 4.1$, $a'=1$ and
$Var\left(p_T\right)/\langle p_T\rangle = 0.43$. We stress here
this fact because keeping the relevant independent parameters
(altogether $4$ of them) as free, unconstrained by the
multiplicity fluctuations data, would result in a simple formula
which could fit data more exactly,
\begin{eqnarray}
\!\! \frac{\sqrt{Cov\left(p_{Ti},p_{Tj}\right)}}{\langle
p_T\rangle}\!\! &=&\!\! \sqrt{\frac{c_1 + c_2\left( 1\! -\!
\frac{N_P}{A}\right)}{c_3N_P+c_4\left( 1\! -\!
\frac{N_P}{A}\right)}}.
\end{eqnarray}
To close let us stress again that our approach, which uses the
$q$-statistics, allows to demonstrate the deep connection between
fluctuations and correlations observed in transverse momenta and
those observed in multiplicity distributions. It means that they
all convey essentially the same information on the hadronizing
system produced in high energy heavy ion collisions.\\

GW thanks the Organizers of the ISMD2009 for their warm
hospitality during the conference.


\begin{thebibliography}{99}

\bibitem{T} {\it Topical Issue on Statistical Power Law Tails in
            High-Energy Phenomena}, Ed. Bir\'{o} T S (2009)
            Eur. Phys. J. A 40(3).

\bibitem{EPJA} Wilk G, and W\l odarczyk Z (2009) Eur. Phys. J. A 40:
               299 - 312.


\bibitem{OurPaper} Wilk G, W\l odarczyk Z (2009) Phys. Rev C 79:
                   054903(10pp)

\bibitem{Lin} Lindhard J, {\it 'Complementarity' between energy and temperature},
              in {\it The Lesson of Quantum Theory},  Eds. de Boer J, Dal E,
              Ulfbeck O (1986) (North-Holland, Amsterdam).

\bibitem{UnRel} Uffink J, and van Lith J (1999) Foundations of
                Phys. 29: 655 and {\it Thermodynamic
                uncertainity relations}, cond-mat/9806102.

\bibitem{limq}  Campisi M (2007) Phys. Lett.  A 366: 335.

\bibitem{STAR} Adams J et al. (STAR Collab.) (2005)  Phys. Rev. C 72:
               044902 (6pp).

\bibitem{OurOld} Utyuzh O V, Wilk G, W\l odarczyk Z (2001) Phys.
                 Rev. C 64: 027901 (4pp).

\end{thebibliography}
\end{document}